# Condensation and Jumping Relay of Droplets on Lotus Leaf

Cunjing Lv, Pengfei Hao,[*] Zhaohui Yao, Yu Song, Xiwen Zhang, Feng He

Department of Engineering Mechanics, Tsinghua University, Beijing 100084, China

**Abstract**

Dynamic behavior of micro water droplet condensed on a lotus leaf with two-tier roughness is studied. Under laboratory environment, the contact angle of the micro droplet on single micro papilla increases smoothly from 80° to 160° during the growth of condensed water. The best-known "self-clean" phenomenon, will be lost. A striking observation is the out-of-plane jumping relay of condensed droplets triggered by falling droplets, as well as its sustained speed obtained in continuous jumping relays, enhance the automatic removal of dropwise condensation without the help from any external force. The surface tension energy dissipation is the main reason controlling the critical size of jumping droplet and its onset velocity of rebounding.

Vapor condensation is widely observed in nature and plays an essential role in energy conversion, water harvesting, thermal management, and ballistospore discharge [1-8]. Because of its fundamental importance, the physical process of water condensation on wetting behaviors of natural and artificial superhydrophobic surfaces has attracted wide attention in recent years [9-12]. Previous studies have mainly focused on the size change and the movement of the droplet [13-15], while variation of droplet contact angle on superhydrophobic surface during condensation has rarely been studied. It should be noted that most of the microscopic measurement has been carried out by environmental scanning electron microscopy (ESEM) in low-pressure and low-vacuum environments [16-19], which are different from the actual condition in nature. Our fundamental understanding of the wetting state of droplet condensation on the micro/nano structures remains extremely limited. Accordingly, in this work, we present the twofold unexpected and spectacular wetting phenomena under laboratory environment: (1) in strong contrast to drop a water droplet on lotus leaf, the famous phenomenon, "self-clean effect", is lost, during the growth of micro droplets on single micro-papilla with two-tier roughness, the time dependence of the contact angle and droplet radius is reported and systematically investigated; (2) out-of-plane jumping relay of condensed droplets triggered by falling droplet, for the first time, is observed, and the continuous jumping relay could make the droplets obtain sustained speed and enhance the automatic removal of dropwise condensation without any external force. We have also developed general models to explain these results, which accounts quantitatively for determining jumping velocities of coalesced condensed droplets.

---

[*] To whom correspondence should be addressed. E-mail: haopf@tsinghua.edu.cn

Live and fresh lotus leaves were procured from a pond on campus of Tsinghua University, Beijing, China. ESEM images in Fig. 1 shows that the lotus leaf surface is composed of a large number of micro-papilla with average diameter of $10\mu m$. Interestingly, larger papillae with diameter of $40 \sim 60\mu m$ were observed among the small papillae, which were not mentioned in previous works [13,16]. Fig. 1(b) also shows that both large and small papillae are covered with countless nano-hairs. Samples of the lotus leaves ($15mm \times 1mm$) were mounted horizontally on a peltier cooling stage using double sided adhesive tape and the temperature of the samples was maintained at $5°C \pm 1°C$. The environmental temperature and the relative humidity were maintained well at $28°C \pm 1°C$ and $44 \pm 2\%$, respectively. The apparent contact angle (ACA) of the droplet on the sample was measured with a commercial contact angle meter (JC2000CD1). Side view images of water condensing from the vapor were captured by a high-speed camera (200fps, $640 \times 480 pixel$, PIKE-F032, Germany) with $1.5\mu m \times 1.5\mu m$ spatial resolution. Compared with observation by employing ESEM or optical microscope, this method is more convenient to obtain the base diameter and ACA of the micro-droplet on the lotus leaf with high image acquisition rate.

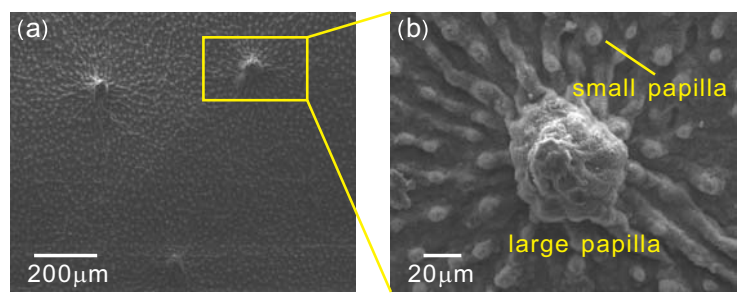

FIG. 1. SEM images of a lotus leaf in different scale: (a) small and large micro-papillae distributed on the lotus leaf; (b) enlargement of the micro-papillae.

The condensation process of a droplet in micro-scale on the top of a single large micro-papilla of lotus leaf can be identified as two stages: In the first stage, plenty of nucleated nanodroplets are formed and fill the valleys among the nanostructures on the top of the papilla, then a microdroplet is formed by the coalescence of these gradually growing nanodroplets. As shown in the first 40s in Fig. 2(a)(b)(c) and Fig. 3(a), the microdroplet on the top of the micro-papilla almost kept its ACA by increasing the base diameter and it was in the constant contact angle (CCA) wetting state. The low value of ACA ($<90°$) indicates that the droplet might be in a fully Wenzel wetting state [20] or a partially Cassie-Baxter (CB) wetting state [21], which is different from a sessile droplet on the lotus surface in CB wetting state [22], but consistent with the ESEM observation of the droplet condensation on the natural or artificial superhydrophobic surfaces with two-tier roughness or nanostructures [23,24]. In the second stage, as shown in Fig. 2(d)(e), when the base diameter of the microdroplet reaches the size of the top spot of the papilla, contact line pins and the constant contact line (CCL) wetting state is observed. During its further growth, the ACA of the droplet increases smoothly from $\sim 100°$ to $\sim 160°$. However, the condensed droplet will not roll off with such large ACA even if the sample is vertically oriented, which verifies the possible explanation that the condensation liquid fully or partially fills the air gap among the nano-haris of the lotus leaf, leading the contact area and the adhesion force between the solid and liquid increase. The famous "self-clean effect" of lotus leaves will be totally lost for condensed water droplet, and the reason can be attributed to the process how droplets are formed.

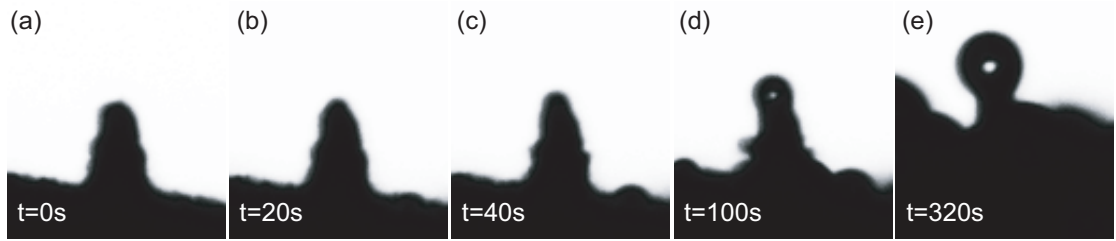

FIG. 2. Time sequence images of the growth of an isolated droplet on the top of single micro-papilla. In the first 30s the droplet was in a CCA wetting state but subsequently the boundary contact line was pinned and CCL wetting states happened.

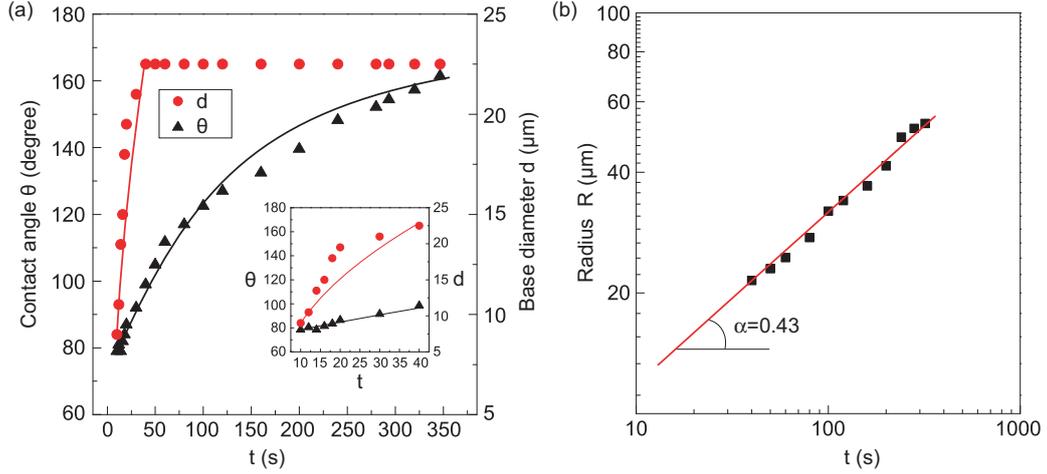

FIG. 3. (a) Time evolution of the base diameter (red solid circles) and the ACA (black triangles) of an isolated droplet in the large micro-papillae of the lotus leaf. The insert is the time evolution in early stages. The solid red and black lines are used to guide the eye [25]. (b) Time dependence of the droplet radius after condensation on a single micro-papilla obeys a power law $R \sim t^\alpha$, here $\alpha \approx 0.43$.

On the CCL wetting state, the time dependence of the droplet radius condensed on single micro-papilla obeys a power law [13,26-28]:

$$R \sim t^\alpha. \qquad (1)$$

Due to limitations of the high-speed camera, it is difficult to measure the radius accurately and determine the power law in the first 40s. As shown in Fig. 3, the growth followed a power law with $a \approx 0.43$, which is close to the growth rates observed on hydrophobic surface [26], the deviation from the 1/3 law may come from the pinning of the contact line [26-30].

Self-propelled dropwise condensation was reported by Boreyko on superhydrophobic surfaces [13]. To the best of our knowledge, there is no report about self-propelled droplet on plant leaves. Coalescence-triggered self-propelled dropwise condensation was observed on the lotus leaf by using a high-speed camera (200fps) with long exposure time (4ms). Fig. 4(a) records the jump trajectory of a droplet with $\sim 10\mu m$ diameter. The initial jump velocity is $\sim 0.06\,m/s$, which is consistent with the experimental results carried on the artificial superhydrophobic substrate with two-tier roughness [13]. Herein, both the dropwise condensation and self-propelled single droplet on the micro-papilla of the lotus leaf are similar to the growth of the Buller's droplet [4,5].

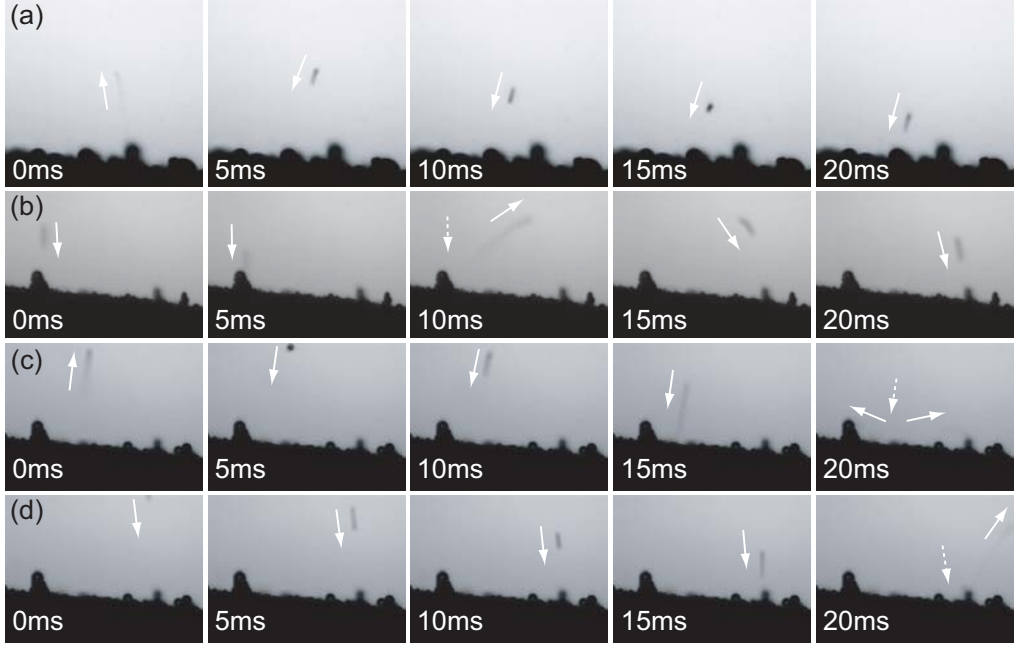

FIG. 4. Various bouncing behaviors of coalescence-induced self-propelled droplet on lotus leaf: (a) The droplet first reaches the highest point and then falls. (b) Droplet rebounds at higher velocity, triggered by a falling droplet. (c) After one bounding droplet falls and touches the lotus leaf, it triggers two droplets to bounce. (d) After a bounding droplet falls and touches the lotus leaf, it rebounds twice: on the lotus leaf and on the top of a micro-papilla, respectively. The dashed lines are the traces of the previous falling droplets.

In strong contrast to the reported self-propelled dropwise condensate on superhydrophobic surfaces [6,8,13,15,17,18], a distinguishing phenomenon is that there are continuous jumping relays happening on the droplets with sustained speed, which remarkably enhances the automatic removal of dropwise condensation without the help from any external force, the continuous jumping relays observed in our experiments has not been reported in the past. Fig. 4(b)(c)(d) illustrates the selected high speed camera snapshots of droplet jumping relays thanks to the long exposure time. Interestingly, we demonstrate here that compared with the falling velocity ($\sim 0.01\,\mathrm{m/s}$) when the droplet just touched the lotus leaf, the onset of the rebounding velocity ($\sim 0.12\,\mathrm{m/s}$) is much higher. It indicates that this jumping relay is triggered not by rebounding [30], but by the falling droplet coalescing with other droplets and its kinetic energy is obtained from the change of surface tension during droplet coalescence on the lotus leaf. As shown in Fig. 4(b)(c)(d), in a more perspective, the behavior of the droplet jumping relay is endowed with the follow characters: random, fast motion, happens in a wide area, multiple relays, one droplet could trigger multiple droplets. The continuous jumping relays observed in this letter may find a wide range of applications where high efficient removal of condensed water droplet is needed, such as thermal management, anti-fogging as well as anti-icing.

Even though researchers developed various models [13,32,33] to predict rebound velocity of droplet on superhydrophobic surfaces very recently, neither formulae includes the influence of the nano-texture of two-tier structures on the self-propelled motion of a coalesced condensed droplet, so velocity predictions in those models deviate from the experimental observations and the underlying physical meanings has not been revealed very well yet. For a droplet on a substrate with certain apparent contact angle $\theta$, its whole potential energy could be expressed as $\Pi = E_{\text{surf}} - \Delta p \cdot V_0$ with the increase of surface tension energy $E_{\text{surf}} = \gamma_{\text{LV}} A_{\text{LV}} + (\gamma_{\text{SL}} - \gamma_{\text{SV}}) A_{\text{SL}} = 3\gamma_{\text{LV}} V_0 / r$ and also dissipated with the change of droplet morphology $\Delta p \cdot V_0$, here, $\gamma_{\text{LV}}$, $\gamma_{\text{SL}}$ and $\gamma_{\text{SV}}$ are the surface tensions at the solid/liquid/vapor interface, and the capillary pressure $\Delta p$ is defined as $\Delta p = p_{\text{in}} - p_{\text{out}} = 2\gamma_{\text{LV}} / r$ [34]. Accordingly, we can get the change of the total energy about the two droplets before and after coalescence,

$$\Delta \Pi = 2V_0 \gamma_{\text{LV}} \left( \frac{1}{r_1} - \frac{1}{r_2} \right) \tag{2}$$

where, $r_1$ is the radius of each droplet with volume $V_0$ before coalescence and $r_2$ is the radius of the droplet with volume $2V_0$ after coalescence, which could be determined by the apparent contact angle $\theta_1$ and $\theta_2$, respectively.

During droplet coalescence, the energy loss in deforming the droplet against viscosity is $\Phi = \boldsymbol{\tau} : \nabla \mathbf{u}$ [35], where $\boldsymbol{\tau}$ and $\mathbf{u}$ are the shear stress and velocity inside the droplet, $\nabla$ is the gradient operator in three-dimensional space. Different from Wang's theory [29], it's more plausible to believe that the dissipation caused by the viscosity mainly comes from the moving direction (x-direction, for example) of two droplets merging with each other, so we can give a more appropriate estimation $\Phi = 2\tau_{xx}\tau_{x,x} \approx \mu(u/r_1)^2/2$ for each droplet. By employing the characteristic capillary time scale $t \propto \sqrt{\rho r_1^3 / \gamma_{\text{LV}}}$ [13] and $\Delta p$, an estimation of the average coalesced velocity could be obtained as $u \approx t \cdot \Delta p \cdot A_{\text{sec}} / \rho V_0 \approx (3/2)(A_{\text{sec}}/A_0)\sqrt{\gamma_{\text{LV}}/\rho r_1}$. So, we can get the dissipation for each droplet,

$$\Delta E_{\text{vis}} = \int_0^t \int_{V'} \Phi dV' dt' \approx \frac{3}{2} \mu \pi \left( \frac{A_{\text{sec}}}{A_0} \right)^2 \sqrt{\frac{\gamma_{\text{LV}} r_1^3}{\rho}} \tag{3}$$

where $A_0 = \pi r_1^2$, and $A_{\text{sec}} = r_1^2 (\theta_1 + \sin\theta_1 \cos\theta_1)$ is the cross section of a single droplet.

It's well known that whether coalescence induced jumping could happen or not mainly depends on the wetting characters of the substrate, because the coalescence offers a mechanism to switch from the sticky Wenzel state [20] to the non-sticking Cassie state [21], the energy barrier must be overcame by the pinning of the three-phase contact lines [29,30]. Surface tension energy dissipation caused by contact angle hysteresis could be estimated by the Young-Dupré Equation [34],

$$\Delta E_{\text{hys}} = A_{\text{cont}} \gamma_{\text{LV}} (1 + \cos\theta_Y) \tag{4}$$

Here, $\theta_Y$ is the intrinsic contact angle of the materials on lotus leaf. The contact area of the two droplets on the lotus leaf before coalescence is defined as $A_{\text{cont}} = 2\pi r_f (r_1 \sin\theta_1)^2$, where $r_f$ is the roughness of the substrate. Based on the law of conservation of energy, we can give an estimation of the onset velocity of the jumping relay,

$$\rho V_0 U^2 = 2V_0 \gamma_{\text{LV}} \left(\frac{1}{r_1} - \frac{1}{r_2}\right) - A_{\text{cont}} \gamma_{\text{LV}} (1 + \cos\theta_Y) - 3\mu\pi \left(\frac{A_{\text{sec}}}{A_0}\right)^2 \sqrt{\frac{\gamma_{\text{LV}} r_1^3}{\rho}} \tag{5}$$

where $U$ is the droplet velocity after coalescence happens. The contribution of the gravity potential in this case is so small that it is ignored in Eq. (5), as well as the kinetic energy of the droplet before coalescence. As shown in Fig. 5, our theoretical prediction is very consistent with the experimental measurement.

By comparing of each energy contribution and dimensional analysis, Eq. (5) will lead to a general formulae $U^2 \sim \gamma_{\text{LV}} (\rho r_1)^{-1} \left[c_1 - c_2 \mu (\gamma_{\text{LV}} \rho)^{-1/2} r_1^{-1/2}\right]$, where, $c_1$ and $c_2$ are dimensionless physical coefficients (see Supplemental Information). Interestingly, it demonstrates that the first term is the net surface tension energy during the condensation which characteristics the critical velocity $U_c \propto \sqrt{\gamma_{\text{LV}}/\rho r_1}$, and the second term characteristics the critical size of the rebounding droplet $r_c \propto \mu^2/\gamma_{\text{LV}} \rho$ [12]. But in the real case, the previous theoretical prediction of the critical diameter for jumping is much lower than what is experimentally observed [13,32,33]. Because of the dispassion of the surface tension energy during coalescence, $c_1$ is in the order of $10^{-2} \sim 10^{-1}$ (see Supplemental Information), much lower than a unit, which makes the critical size of the droplet as $10^2 \sim 10^4$ larger than $\mu^2/\gamma_{\text{LV}} \rho$, which is why the experimentally observed critical diameter for jumping was of order 10μm [8,13,15,24]. To a certain extent, the net surface tension energy is determined by the contact angle hysteresis and in turn limit the onset velocity and the critical size of the droplet rebounding.

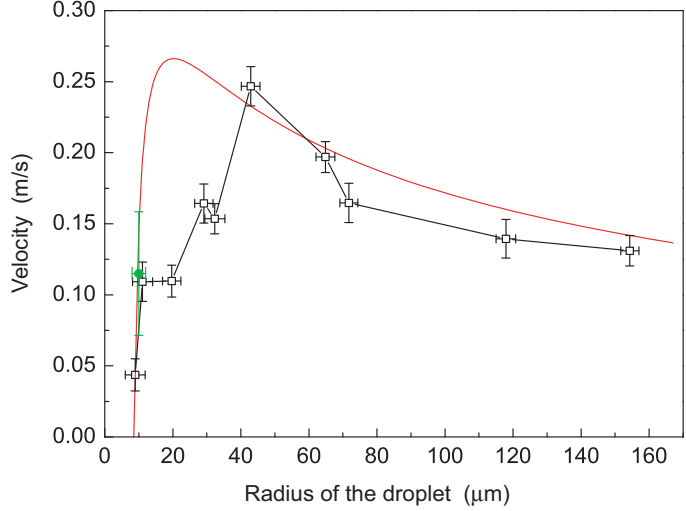

Fig. 5. Comparison between experimental results and theoretical prediction of droplet velocity as a function of droplet radius. The black hollow squares are Boreyko's results [9]; the green solid circle are five average value of our experiment as shown in Fig. 4(a); the solid red line is from predictions by Eq. (7) with $\theta_1^*=151°$, $\theta_2^*=170°$, $\theta_Y=105°$, $r_f=1.36$.

In conclusion, both CCA and CCL wetting states on the growth of water droplet on the micro-papilla of lotus leaf under natural environment, as well as the remarkable out-of-plane continuous jumping relay of condensed droplet trigged by falling droplet, were first reported in this letter. There is still a challenge to remove sticky small water droplets even on superhydrophobic materials. Interestingly, the remarkable continuous jumping relay behavior will lead a sustained speed ballistic movement in every relay and greatly enhance the automatic removal of dropwise condensate without the action of any external force. We further show that the surface tension energy dissipation during droplet transition and contact angle hysteresis limits the size of the jumping droplet in micro-scale. We believe the discovery of the special jumping relay of droplets on lotus leaf, coupled with the fundamental theory, provide insights into the liquid-solid interactions at the micro/nano-scale and may lead to promising applications including high heat flux applications, fast dropwise condensation, and anti-fouling.

**Acknowledgements:** Financial support from the NSFC under grant Nos.11072126, 11272176 and 11172156 is gratefully acknowledged.

# Supplementary Information for

## Condensation and Jumping Relay of Droplets on Lotus Leaf

by Cunjing Lv, Pengfei Hao,[*] Zhaohui Yao, Yu Song, Xiwen Zhang, Feng He

[*]To whom correspondence should be addressed. E-mail: haopf@tsinghua.edu.cn

**S.1 Theory - Dimensional analysis about the contribution of the energy**

Based on the quantitative estimation of Eq. (5),

$$\rho V_0 U^2 = 2V_0 \gamma_{LV}\left(\frac{1}{r_1} - \frac{1}{r_2}\right) - A_{cont} \gamma_{LV}(1+\cos\theta_Y) - 3\mu\pi\left(\frac{A_{sec}}{A_0}\right)^2 \sqrt{\frac{\gamma_{LV} r_1^3}{\rho}} \quad (5)$$

For convenience, we can define the droplet volume $V_0$ by employing different geometrical parameters,

$$\begin{aligned} V_0 &= \frac{4}{3}\pi r_s^3 \\ &= \frac{1}{3}\pi r_1^3 (1-\cos\theta_1)^2 (2+\cos\theta_1) \\ &= \frac{1}{6}\pi r_2^3 (1-\cos\theta_2)^2 (2+\cos\theta_2) \end{aligned} \quad (S1)$$

Denoting by $r_s$ the radius of a spherical droplet with volume $V_0$, $r_1$ and $r_2$ are the radii of the droplets before and after coalescence. According to Eq. (S1), it's not difficult to give an explicit expression for Eq. (5),

$$\begin{aligned} U^2 &= \frac{2\gamma_{LV}}{\rho r_1}\left[1 - \left(\frac{r_1}{r_2}\right) - \frac{3}{4}\left(\frac{r_1}{r_s}\right)^3 r_f \sin^2\theta_1 (1+\cos\theta_Y) - \frac{9}{8}\left(\frac{A_{sec}}{A_0}\right)^2 \left(\frac{r_1}{r_s}\right)^3 \frac{\mu}{\sqrt{\gamma_{LV}\rho}} \cdot \frac{1}{\sqrt{r_1}}\right] \\ &= \gamma_{LV}(\rho r_1)^{-1}\left[c_1 - c_2 \mu(\gamma_{LV}\rho)^{-1/2} r_1^{-1/2}\right] \end{aligned}$$

(S2)

with

$$c_1 = 2 - 2\left(\frac{r_1}{r_2}\right) - \frac{3}{2}\left(\frac{r_1}{r_s}\right)^3 r_f \sin^2\theta_1 (1+\cos\theta_Y) \quad (S3)$$

$$c_2 = \frac{9}{4}\left(\frac{A_{sec}}{A_0}\right)^2 \left(\frac{r_1}{r_s}\right)^3. \quad (S4)$$

As we know, the prerequisite for droplet jumping on superhydrophobic substrates is the high value of apparent contact angles ($\theta_1$ and $\theta_2$) before and after coalescence. Here, we want to emphasize that without loss of generality, we assume $\delta_1 = \pi - \theta_1 \ll 1$ and $\delta_2 = \pi - \theta_2 \ll 1$ in order to give dimensional analysis depending on Taylor series expansions. We can get,

$$\frac{r_1}{r_s} = \sqrt[3]{\frac{4}{(1+\cos\delta_1)^2(2-\cos\delta_1)}} \approx 1 + \frac{1}{16}\delta_1^4 + O(\delta_1^6) \tag{S5}$$

$$\frac{r_2}{r_s} = \sqrt[3]{\frac{8}{(1+\cos\delta_2)^2(2-\cos\delta_2)}} \approx 2^{1/3}\left[1 + \frac{1}{16}\delta_2^4 + O(\delta_2^6)\right]. \tag{S6}$$

Put Eq. (S5) and Eq. (S6) to Eq. (S2)-(S4), the order of each geometrical parameter could be estimated,

$$c_1 = 2 - 2\left(\frac{r_1}{r_2}\right) - \frac{3}{2}\left(\frac{r_1}{r_s}\right)^3 r_f \sin^2\theta_1 (1+\cos\theta_Y) \tag{S7}$$
$$\approx 0.4126 - 1.5 \cdot r_f \sin^2\theta_1 (1+\cos\theta_Y)$$

$$c_2 \sim 10^0. \tag{S8}$$

On lotus leaf, we have $\theta_Y \approx 105°$ and $r_f > 1$. As mentioned above, for jumping droplet in superhydrophobic wetting sate, it is reasonable to assume $\theta_1 \geq 150°$. On the basis of the above analysis, we can conclude that the value of the parameter $c_1$ must be in the range of 0.01 to 0.1, so, the critical size of the jumping droplet is,

$$r_1 = \left(\frac{c_2}{c_1}\right)^2 \cdot \frac{\mu^2}{\gamma_{LV}\rho} \approx \left(\frac{c_2}{c_1}\right)^2 \cdot 14\,\text{nm} \sim 10^1\,\mu\text{m}, \tag{S9}$$

which is critical important to understanding droplet jumping phenomena for condensed droplet on surperhydrophobic materials. From the above analysis, the influence of contact angle hysteresis, say $r_f \sin^2\theta_1 (1+\cos\theta_Y)$, subtly determine the net surface tension energy, accordingly, limit the onset velocity and the critical size of the droplet rebounding.

**S.2 Supplementary Figure**

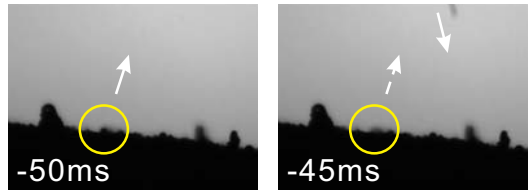

FIG. S1. Detailed information for Fig. 4(b): enlargement of the onset of the coalescence-induced self-propelled droplet bounce (see also Video S1). The dashed line is the trace of the bouncing droplets.

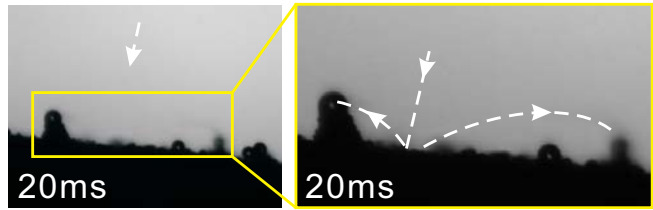

FIG. S2. Detailed information for Fig. 4(c): enlargement of the moment in which two droplets are triggered to bounce by one falling droplet. The dashed lines are the traces of the falling and bouncing droplets.

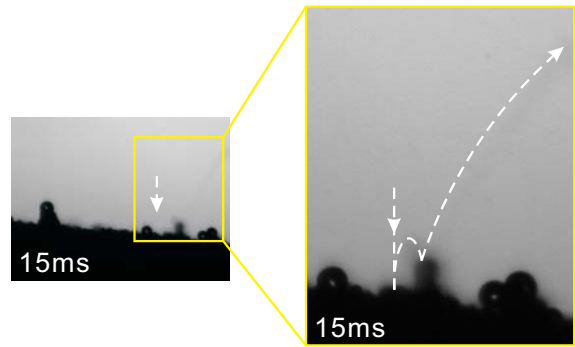

FIG. S3. Detailed information for Fig. 4(d): enlargement of the twice of the rebounding droplet on the lotus leaf and on the top of a micro-papilla, respectively. The dotted lines are the traces of the falling and bouncing droplets.

**S.3 Supplementary Video**

**Description:**

**Video S1 (S1.mpg, 248KB, for Fig 4(b), Fig. S1).** Side-view imaging of the jumping behaviors of coalesced droplet: firstly, jumping of the droplet is caused by coalescence. After it reaches the highest point, it will be in free falling until it touches a stationary droplet on the lotus leaf, the stationary droplet can be triggered by this falling droplet and the merged droplet will obtain sustained speed in this relay. The video was captured at 200 fps and played back at 4fps.

**Video S2 (S2.mpg, 98KB, for Fig 4(c), Fig. S2).** Side-view imaging of a jumping

coalesced droplet: after the droplet touches the louts leaf, other two droplets are triggered by this droplet. The video was captured at 200 fps and played back at 4fps.

**Video S3 (S3.mpg, 82KB, for Fig 4(d), Fig. S3).** Side-view imaging of a jumping coalesced droplet: After a bounding droplet touches the lotus leaf, it rebounds twice on the lotus leaf and on the top of a micro-papilla, respectively. The video was captured at 200 fps and played back at 4fps.